\documentclass[conference]{IEEEtran}
\IEEEoverridecommandlockouts
\usepackage{cite}
\usepackage{amsmath,amssymb,amsfonts}
\usepackage{algorithmic}
\usepackage{graphicx}
\usepackage{textcomp}
\usepackage{xcolor}
\usepackage{amsmath}
\usepackage{amssymb}
\usepackage{breqn}
\usepackage {hyperref}
\usepackage[utf8]{inputenc}
\usepackage{float}

\newcommand{\etal}{\textit{et al}.}
\newcommand{\ie}{\textit{i}.\textit{e}.}

\usepackage{fancyhdr, lipsum}
\fancypagestyle{firstpage}{
   \fancyhf{} 
   \newcommand{\changefont}{%
    \fontsize{9}{11}\selectfont
    }
    \headheight 40pt

   \cfoot{\thepage}
   \fancyhead[L]{\changefont © 20XX IEEE.  Personal use of this material is permitted.  Permission from IEEE must be obtained for all other uses, in any current or future media, including reprinting/republishing this material for advertising or promotional purposes, creating new collective works, for resale or redistribution to servers or lists, or reuse of any copyrighted component of this work in other works.}
}
\def\BibTeX{{\rm B\kern-.05em{\sc i\kern-.025em b}\kern-.08em
    T\kern-.1667em\lower.7ex\hbox{E}\kern-.125emX}}
\begin{document}

\title{Dual Attention Network for Heart Rate and Respiratory Rate Estimation\\
{
}
}

\makeatletter

\def\ps@IEEEtitlepagestyle{%
  \def\@oddfoot{\mycopyrightnotice}%
  \def\@evenfoot{}%
}
\def\mycopyrightnotice{%
  {\footnotesize
  \text{978-1-6654-3288-7/21/\$31.00\textcopyright 2021IEEE}
  }
  \gdef\mycopyrightnotice{}
}

\hyphenation{op-tical net-works semi-conduc-tor}


\makeatother


\iftrue 
\author{
   \IEEEauthorblockN{
        Yuzhuo Ren\IEEEauthorrefmark{1}, Braeden Syrnyk\IEEEauthorrefmark{1}\IEEEauthorrefmark{2}, Niranjan Avadhanam\IEEEauthorrefmark{1}
    }
    \IEEEauthorblockA{\IEEEauthorrefmark{1} \textit{NVIDIA}, Santa Clara, USA \\
 }
    \IEEEauthorblockA{\IEEEauthorrefmark{2} \textit{University of Waterloo}, Ontario, Canada\\
    Email: yren@nvidia.com, bsyrnyk@nvidia.com, navadhanam@nvidia.com
    }
}
\fi

\maketitle

\thispagestyle{firstpage}

\begin{abstract}
Heart rate and respiratory rate measurement is a vital step for diagnosing many diseases. Non-contact camera based physiological measurement is more accessible and convenient in Telehealth nowadays than contact instruments such as fingertip oximeters since non-contact methods reduce risk of infection. However, remote physiological signal measurement is challenging due to environment illumination variations, head motion, facial expression, etc. It's also desirable to have a unified network which could estimate both heart rate and respiratory rate to reduce system complexity and latency. We propose a convolutional neural network which leverages spatial attention and channel attention, which we call it dual attention network (DAN) to jointly estimate heart rate and respiratory rate with camera video as input. Extensive experiments demonstrate that our proposed system significantly improves heart rate and respiratory rate measurement accuracy. 
\end{abstract}


\section{Introduction}
Non-contact camera-based physiological measurement is a fast growing research field and draws significant attention especially during the COVID-19 pandemic. Non-contact camera-based physiological measurement reduces infection risks and enables Telehealth, remote health monitoring and smart hospitals \cite{smart-hospital}. The underlying principle for camera-based physiological measurement is capturing subtle skin color changes \cite{wu2012eulerian} or subtle motions \cite{balakrishnan2013detecting} caused by blood circulation. Camera-based physiological measurement involves capturing subtle changes from the body caused by light reflection. Imaging techniques can be used to measure volumetric changes of blood in the surface of the skin by capturing subtle skin color and motion changes. Imaging Photoplethysmography (iPPG) technology is based on the measurement of subtle changes in light reflected from the skin. Image Ballistocardiogram (iBCG) technology is based on the measurement of mechanical force of blood pumping around the body which causes subtle motions. Both heart rate and respiratory rate can be recovered using iPPG or iBCG based methods \cite{bartula2013camera, chen2018deepphys, janssen2015video, wang2015novel}. Camera-based heart rate and respiratory rate estimation is challenging because the skin color change and motions caused by blood circulation is so subtle that it's easily corrupted by environment illumination variations, head motion, facial expression, etc. 
\begin{figure}[htbp]
\includegraphics[height=6cm]{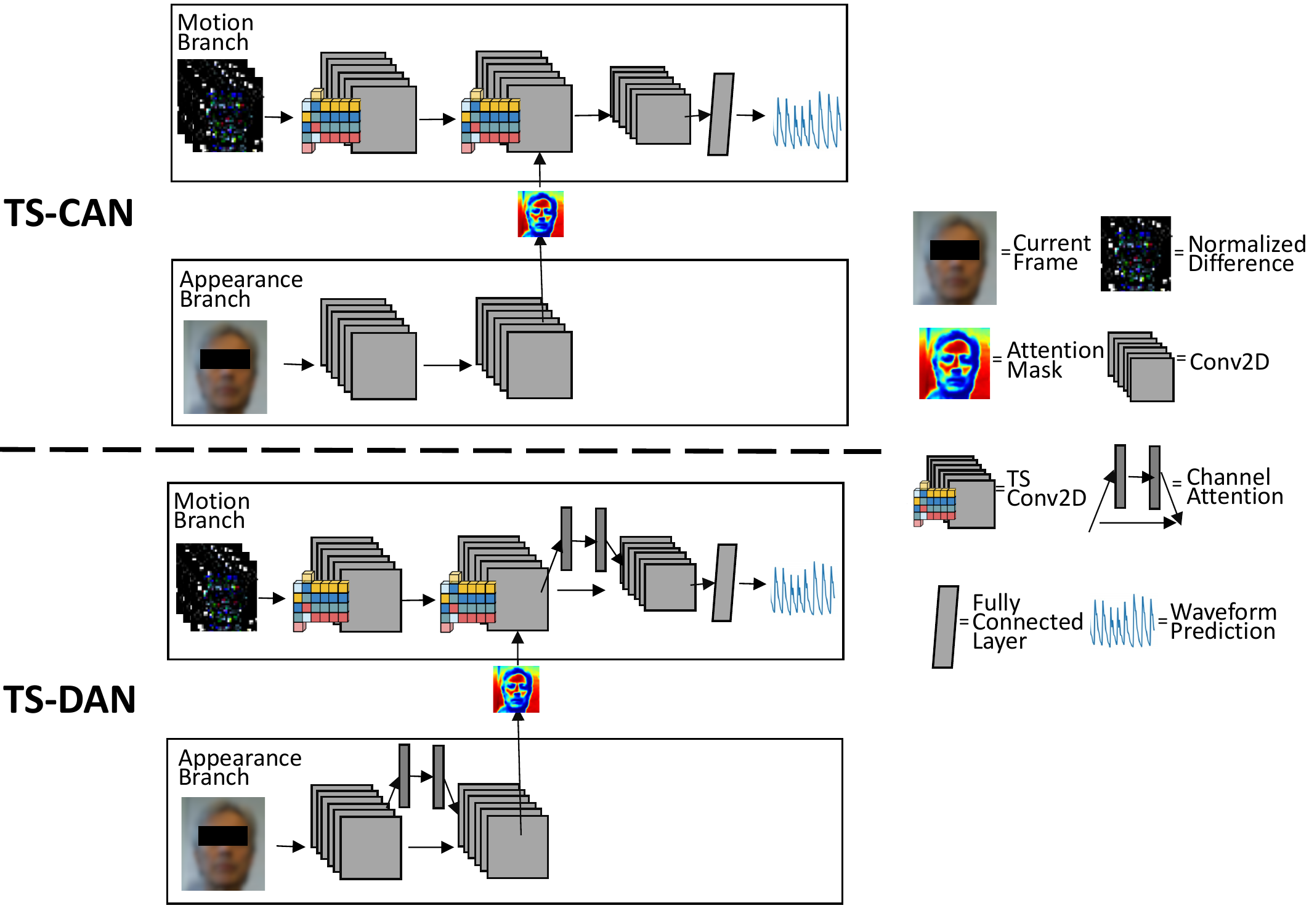}
\caption{Overview of our proposed temporal shift dual attention network (TS-DAN) for estimating heart rate and respiratory rate compared to previous proposed temporal shift convolutional attention network (TS-CAN) \cite{liu2020multi}. TS-DAN leverages both spatial attention and channel-wise attention. Each of these models can be applied in a single or multi-task fashion.}
\label{flow}
\end{figure}

Traditional computer vision based heart rate and respiratory rate estimation involves several components in order to get robust measurements, such as face tracking \cite{wang2016algorithmic}, skin segmentation \cite{tasli2014remote}, heart rate or respiratory rate frequency band filtering\cite{wang2017robust}, principle component analysis (PCA) \cite{lewandowska2011measuring}, etc. Various algorithms have been proposed to improve robustness under each challenging scenarios, such as environment illumination change, head motion, facial expression, etc. Recently proposed convolutional neural networks enable end to end learning of the heart rate and respiratory rate and leverages big data which greatly outperforms traditional hand-crafted feature based methods especially for challenging cases. Several recent approaches have shown the benefit of enhancing spatial representation via spatial attention and channel-wise representation via channel-wise attention to boost the representational power of a convolutional neural network in various research fields. \figurename~\ref{flow} shows our proposed temporal shift dual attention network (TS-DAN) for estimating heart rate and respiratory rate compared to previously proposed temporal shift convolutional attention network (TS-CAN). Each of the networks can estimate heart rate and respiratory rate individually or jointly estimate in a multitask fashion. In our work, we leverage both spatial attention and channel-wise attention to improve convolutional neural network for heart rate and respiratory rate estimation.

\begin{figure*}[ht]
\centering
\includegraphics[height=5.3cm]{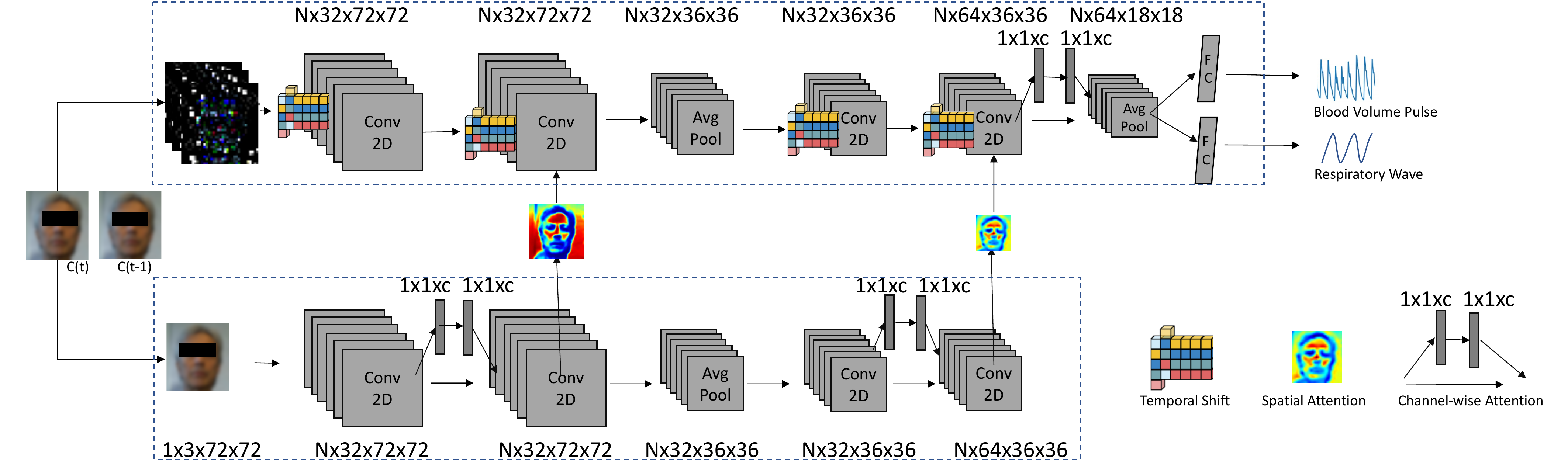}
\caption{We present a multi-task temporal shift dual attention convolutional network (MT-TS-DAN) for joint heart rate and breath rate measurement.}
\label{architecture}
\end{figure*}

We summarize of our contributions as follows: We apply attention mechanism in both spatial domain and channel-wise domain to improve network accuracy. Spatial domain attention enhances spatial encoding which locates facial regions that contain strong physiological signal response. Channel-wise domain attention recalibrates channel-wise feature responses to select the most informative features. To our best knowledge, our proposed network is the first one leveraging both spatial attention and channel-wise attention, existing networks for heart rate and respiratory rate estimation do not have both spatial attention and channel-wise attention in the network architecture.

\section{Related Work}
\label{sec:related_work}

\textbf{Traditional Hand-crafted Feature Method.} Haan \textit{et al.} \cite{de2013robust} proposed CHROM method which leverages light absorption differences among R, G and B channels to conduct noise reduction. Wang \textit{et al.} \cite{wang2015novel} improved motion robustness of CHROM method by using spatially redundant pixel-sensors of a camera, and leveraged artifacts as additional input channels to discriminate pulse and distortions \cite{wang2019discriminative}. Wang \textit{et al.} \cite{wang2017robust} proposed sub-band pulse extraction to suppress periodic motions to particularly improve heart rate estimation robustness in fitness scenarios. Lewandowska \textit{et al.} \cite{lewandowska2011measuring} proposed using channel selection and PCA algorithm to separate heart rate signal and noise. RGBIR sensor has also been proposed for physiological signal estimation in order to leverage additional IR channel to improve signal robustness and reduce noise \cite{wang2020modified}.

\textbf{Convolutional Neural Network Method.} Recent CNN based solutions greatly improve robustness of camera-based physiological measurement. Chen \textit{et al.} \cite{chen2018deepphys} proposed a two branch network taking two consecutive frames' face crop difference as motion map and original frame's face crop as appearance map as input with spatial attention mechanism to improve accuracy in head motion cases. Liu \textit{et al.} \cite{liu2020multi} further improved Chen \textit{et al.} \cite{chen2018deepphys}'s network by adding temporal shift module \cite{lin2019temporal} and multitask learning for heart rate and respiratory rate estimation. However, multitask learning decreases heart rate and respiratory rate accuracy because the same network is used to learn both heart rate and respiratory rate. Niu \textit{et al.} \cite{niu2019rhythmnet} proposed to use spatial-temporal map with a deeper backbone network for end to end heart rate estimation. Several CNNs are proposed to estimate heart rate from a highly compressed video \cite{rapczynski2019effects, yu2019remote}.

\textbf{Temporal Module.} 
Since heart rate and respiratory rate is estimated in a time window, temporal process is crucial to improve accuracy than frame based method. 3D convolution based methods was proposed in \cite{liu2020multi} to replace 2D convolution in Chen \textit{et al.} \cite{chen2018deepphys}'s architecture, though it gives better accuracy than 2D convolution module, the complexity is much higher than 2D module. To reduce complexity of 3D convolution, Lin \textit{et al.} \cite{lin2019temporal} proposed temporal shift (TS) module which shifts part of the channels along the temporal dimension to facilitate information exchange among neighborhood frames. TS module achieves the accuracy of 3D CNN and maintains 2D CNN’s complexity as well. It can be inserted into 2D CNNs to achieve temporal modeling. TS module has been widely used in video understanding\cite{lin2019tsm}, gesture recognition \cite{li2019skeleton} and activity recognition \cite{wu2021action}.  

\textbf{Attention Models.}
Spatial attention \cite{jaderberg2015spatial} captures spatial relationship between features and channel-wise attention \cite{hu2018squeeze} recalibrates channel-wise feature responses to select the most informative features. Spatial attention and channel-wise attention modules have greatly improve accuracy in various computer vision tasks, such as image classification \cite{hu2018squeeze}, segmentation \cite{fu2019dual}, etc. Squeeze and Excitation Network (SEN) \cite{hu2018squeeze} significantly improves image classification accuracy by introducing channel-wise attention modules. Fu \textit{et al.} \cite{fu2019dual} appends positional attention and channel attention on top of Fully Convolutional Network (FCN) \cite{long2015fully} to improve image segmentation. Wang \textit{et al.} \cite{wang2020eca} proposed Efficient Channel Attention (ECA) module to improve efficient of SEN. ECA avoids dimensionality reduction and adds cross-channel interaction which preserves performance while significantly decreases model complexity.


\section{Method}
\label{sec:method}
\subsection{Skin Reflection Model}
For the theoretical optical principle of the model, we follow Shafer's dichromatic reflection model (DRM) \cite{wang2016algorithmic} to model lighting reflection and physiological signals. RGB value of the k-th skin pixel in an image can be defined by a time-varying function\cite{chen2018deepphys, liu2020multi}:

\begin{dmath}
\textbf{C}_{k}(t) = \underbrace{I_{0}\cdot(1+\Psi(m(t), \Theta(b(t), r(t))))}_{I(t)} \cdot (\underbrace{\mathbf{u}_{s}\cdot(s_{0} + \Phi(m(t), \Theta(b(t), r(t))))}_{\mathbf{v}_{s}(t)} + \underbrace{\mathbf{u}_{d}\cdot d_{0} + \mathbf{u}_{p}\cdot \Theta(b(t), r(t))}_{\mathbf{v}_{d}(t)}) + \mathbf{v}_{n}(t) \\
\label{reflection}
\end{dmath}

where $\textbf{C}_k(t)$ denotes a vector of the RGB values; $I(t)$ is the illuminace intensity; $\mathbf{v}_{s}(t)$ and $\mathbf{v}_{d}(t)$ are specular and diffusion reflection respectively; $\mathbf{v}_{n}(t)$ denotes camera sensor's quantization noise. $I(t)$, $\mathbf{v}_{s}(t)$ and $\mathbf{v}_{d}(t)$ can all be decomposed into stationary part (\ie, $I_{0}$, $\mathbf{u}_{s}\cdot s_{0}$, $\mathbf{u}_{d} \cdot d_{0}$) and time-varying part (\ie, $I_{0}\cdot\Psi(\cdot)$, $\mathbf{u}_{s}\cdot\Phi(\cdot)$, $\mathbf{u}_{p}\cdot \Theta(\cdot) $) \cite{wang2016algorithmic}, where $m(t)$ denotes all non-physiological variations such as illumination variations from light source, head motion and facial expressions; $\Theta(b(t), r(t))$ denotes time-varying physiological signal which is a combination of both pulse $b(t)$ and respiration $r(t)$ information; $\Psi (\cdot)$ denotes the intensity variation observed by camera; $\Phi(\cdot)$ denotes the varying parts of the specular reflections;  $\mathbf{u}_{s}$ and $\mathbf{u}_{d}$ denotes the unit color vector of the light source and skin-tissue respectively; $\mathbf{u}_{p}$ denotes the relative pulsatile strengths. $I_{0}$ denotes stationary part of illuminance intensity; $s_{0}$ and $d_{0}$ denotes the stationary specular and diffusion reflection respectively.

Skin reflection model in Eq. \ref{reflection} demonstrates that the relation between RGB value of k-th skin pixel $\textbf{C}_{k}(t)$ and physiological signal $\Theta(b(t), r(t))$ is non-linear and the non-linearity complexity can be caused by non-stationary terms, such as illuminance variation, head motion, facial expression, camera intensity variation, etc. A machine learning model is desired to model the complex relationship between $\textbf{C}_{k}(t)$ and $\Theta(b(t), r(t))$.

\subsection{Dual Attention Network Architecture}
\subsubsection{Overview} 

Our proposed temporal shift dual attention multitask network architecture is shown in \figurename~\ref{architecture}. We follow the network architecture proposed in \cite{chen2018deepphys,liu2020multi} which is a two branch network with motion branch and appearance branch. Motion branch takes $N$ consecutive frame's face ROI difference as input. Appearance branch takes current frame's face ROI as input. Temporal shift module \cite{lin2019tsm} is applied before 2D convolution layers in the motion branch. There are two spatial attention layers that get multiplied to motion branch to select informative spatial features. Since physiological signals are not uniformly distributed on human skin, the soft spatial attention mask which gives higher weights in regions where physiological signal are stronger could improve network accuracy. Different from \cite{chen2018deepphys,liu2020multi}, the novel part of our network architecture is that we add channel-wise attention layers to the network architecture. Channel-wise attention is to select discriminative features in channel dimension. Our proposed dual attention network structure can be used to estimate Blood Volume Pulse (BVP) for heart rate estimation or respiratory wave for respiratory rate estimation individually. The network can also estimate both BVP and respiratory wave in a multitask learning fashion.

\subsubsection{Spatial Attention} 
Spatial soft attention mask is generated before average pooling layers using a $1\times1$ convolution filter. Then the attention mask is multiplied with motion branch feature map via element-wise multiplication. The masked feature map $\mathbb{Z}^{k}$ passed to next layer is calculated in Eq. \ref{spatial_attention},

\begin{dmath}
 \mathbb{Z}^{k} = \frac{H_{k}W_{k}\cdot \sigma(\omega^{k} \mathbb{X}_{a}^{k} + b^{k})}{2||\sigma(\omega^{k} \mathbb{X}_{a}^{k} + b^{k})||_{1}} \odot \mathbb{X}_{m}^{k}\\
\label{spatial_attention}
\end{dmath}

where $\sigma(\cdot)$ is sigmoid activation function, $\omega^{k}$ is the $1\times1$ convolution kernel, $b^{k}$ is the bias, $\mathbb{X}_{m}^{k} $ is the motion branch feature map, $\mathbb{X}_{a}^{k} $ is the appearance branch feature map, $\odot$ is element-wise multiplication, $k$ is the layer index,  $H_{k}$ and $W_{k}$ are height and width of feature map.
\subsubsection{Channel-wise Attention}
We insert three channel-wise attention layers into the network: before the 2D convolution extracting attention masks in appearance branch and before final average pooling. By inserting channel-wise attention layer in appearance branch, a better facial attention mask can be generated. By inserting channel-wise attention layer before the final average pooling, it helps the network emphasize informative features and suppress less useful ones. Following Efficient Channel Attention (ECA)\cite{wang2020eca}, in channel attention module, after channel-wise global average pooling, $1D$ convolution is performed followed by a Sigmoid function to learn channel attention. Please refer \cite{wang2020eca} for more details about channel-wise attention module.
\subsubsection{Multitask Learning Loss}
The multitask learning loss is the summation of heart rate pulse waveform MSE loss and respiratory rate waveform MSE loss, which is defined in Eq. ~\ref{multitask_loss},
\begin{dmath}
\textit{L} = \alpha \frac{1}{T} \sum_{t = 1}^{T}(p(t) - p(t)^{'})^{2} + \beta \frac{1}{T} \sum_{t = 1}^{T}(r(t) - r(t)^{'})^{2}, \\
\label{multitask_loss}
\end{dmath}

where $T$ is the time window, $p(t)$ and $r(t)$ are time variant ground truth pulse waveform sequence and respiratory waveform sequence respectively, $p(t)^{'}$ and $r(t)^{'}$ are predicted pulse waveform and respiratory waveform, $\alpha$, $\beta$ are empirical parameters to balance pulse waveform loss, respiratory waveform loss and correlation loss. We set $\alpha=\beta=1$ in our experiment.

\subsubsection{Heart Rate and Respiratory Rate Estimation From Waveform}
The output of our of neural network model is the pulse waveform sequence and respiratory waveform sequence. To extract the heart rate and respiratory rate in beats per minute, a Butterworth bandpass filter was applied to the model outputs with cut-off frequencies of 0.67 and 4 Hz for heart rate, and 0.08 and 0.50 Hz for respiratory rate. The filtered signals were then divided into 10-second windows to apply the Fourier transform to get the dominant frequencies as the heart rate and respiratory rate.

\begin{table*}

\centering
\caption{Ablation on proposed temporal shift dual attention network (TS-DAN) on COHFACE dataset.}
\begin{tabular}{lccc|ccc|ccc}
\hline\noalign{\smallskip}
\bf{Method} & \multicolumn{3}{c}{\bf{Full}}  &  \multicolumn{3}{c}{\bf{Clean}} & \multicolumn{3}{c}{\bf{Natural}}\\
\noalign{\smallskip}
\hline
\noalign{\smallskip}
Heart Rate & MAE & Availability & SNR & MAE & Availability & SNR & MAE & Availability & SNR\\
2D-CAN\cite{chen2018deepphys}   &3.693&0.848&5.479&2.702&  0.846& 5.998&4.683& 0.850& 4.960\\
TS-CAN\cite{liu2020multi}       &1.751&0.907&6.712&1.771&  \textbf{0.963}& 7.099&1.735& 0.850& 6.325 \\
2D-CAN+1ECA                     &2.959&0.867&5.636&1.681& 0.865 & 6.232& 4.237 & 0.869 & 5.040  \\ 
TS-DAN(1ECA)                     &\textbf{1.441}&0.923&6.860&1.150&  0.939& 7.395&\textbf{1.731}& 0.906&6.324  \\
TS-DAN(3ECA)  &1.668&\textbf{0.929}&\textbf{6.939}&\textbf{1.111}&  0.939& \textbf{7.437}&2.224& \textbf{0.919}&\textbf{6.441}    \\
\hline
\end{tabular}
\label{ablation}
\end{table*}

\section{Experiments}
\label{sec:experiments}
We  compare  our  methods  with  two  approaches  for  heart rate  measurement and respiratory rate measurement: 2D convolutional attention network (2D-CAN)\cite{chen2018deepphys} and temporal shift convolutional attention network (TS-CAN) \cite{liu2020multi}. 

\subsection{Datasets}

We run our experiments using the following datasets:

\textbf{COHFACE}\cite{heusch2017reproducible}: The dataset contains RGB videos of faces, synchronized with heart-rate and breathing-rate of the recorded subjects. The video sequences have been recorded with a Logitech HD C525 at a resolution of 640x480 pixels and a frame-rate of 20Hz. Blood volume pulse (BVP) and respiratory rate waveforms are also recorded and synchronized with video timestamp. The dataset includes 160 one-minute long video sequences of 40 subjects (12 females and 28 males). There are 4 videos from every client: 2 videos with good conditions, another 2 videos with more natural conditions. Natural condition videos include ceiling lights OFF and half opened blinds to introduce lightening variations. We follow the training and testing data split protocol provided by COHFACE dataset in our experiment.

\textbf{Our dataset}: The dataset contains RGB videos of face, ground truth heart rate pulse wave recorded from fingertip pulse oximeter. The dataset includes one-minute long video sequence of 15 subjects. Each subject gives 1 session with good condition: similar distance to camera, face static to camera, good illumination condition, and another session with natural condition: subjects seating in various distance to camera ranging from 0.5 meter up to 2 meter, large head motion in roll pitch yaw directions range from 0 to 90 degrees, illuminace variations by adjustable light source.

\subsection{Experiment Details}
We use the COHFACE\cite{heusch2017reproducible} and our dataset to evaluate our proposed convolutional neural network architecture. We use OpenCV face detector to get face crop and resize it to 72x72 for motion map and appearance map generation. Previous work \cite{chen2018deepphys, liu2020multi} resized the face crop to 32x32, however, we found that higher resolution gives better accuracy. Motion map is current frame and previous frame's face crop subtraction. Appearance map is the current frame's face crop. Both motion map and appearance map are normalized in the video.   

To have a fair comparison with previous work 2D-CAN \cite{chen2018deepphys} and TS-CAN\cite{liu2020multi}, we use the same backbone architecture to train 2D-CAN, TS-CAN and our TS-DAN. We also use the same preprocessed motion map and appearance map and the same post processing step to train and evaluate different networks. We evaluate our model in various aspects. First we did ablation on our proposed TS-DAN network architecture, \textit{i.e.}, we compare accuracy by applying ECA \cite{wang2020eca} module in different layers to show how channel-wise attention module could help improve network accuracy. Attention map is visualized and compared with 2D-CAN and TS-CAN. Second, we compare our TS-DAN model accuracy with TS-CAN in single task heart rate and respiratory rate learning to demonstrate our TS-DAN model's accuracy improvement by leveraging both spatial attention and channel-wise attention. Third, we evaluate our proposed TS-DAN network in multitask learning for joint heart rate and respiratory rate estimation. Finally, we evaluate our model's cross-datasets generalization ability, \ie, training our model and previous work model 2D-CAN and TS-CAN using one dataset and test model accuracy on another dataset which collected in different environment settings.

\subsection{Evaluation metrics}
The evaluation metrics were computed over all  windows  of  all  the  test  videos in a dataset, we use the following three metrics:
\begin{itemize}
    \item \textbf{Mean Absolute  Error (MAE):} The average absolute error between ground truth heart/respiratory rate and predicted heart/respiratory rate in beats per minute.
    
    \item \textbf{Signal to Noise Ratio (SNR):} We calculate blood volume pulse and respiration SNR according to the method proposed by De Haan \etal \cite{de2013robust}. The SNR is calculated in the frequency domain as the ratio between the energy around the first two harmonics and remaining frequencies within heart rate and respiratory rate frequency range. SNR captures the quality of predicted heart rate and respiratory rate. $\text{SNR} < 0$ indicates predicted heart/respiratory rate is not reliable since signal energy is less than noise energy. 
    
    \item \textbf{Availability:} We compute the percentage of evaluation window with $\text{SNR} \ge 0$ as availability. This metric captures percentage of the time the system is able to predict high quality heart/respiratory rate. 
\end{itemize}

\section{Results and Discussion}

\subsection{Ablation on Proposed TS-DAN}
The ablation study on our proposed dual attention network is shown in Table \ref{ablation}. We implemented and experimented with multiple combinations of ECA module, TS module, and 2D-CAN model. Model 2D-CAN+1ECA which adds one ECA layer before last average pooling to 2D-CAN already outperforms 2D-CAN even without TS module which demonstrates that ECA module helps select more informative features to improve model accuracy. TS-CAN+1ECA adds TSM module in motion branch (TS module location is shown in \figurename~\ref{architecture}) further improves model accuracy when compared to 2D-CAN+1ECA. TS-CAN+3ECA achieves best accuracy which is a combination of ECA, TSM, and 2D-CAN model where ECA gets applied twice in the appearance branch before each spatial attention module and another ECA gets applied once before the last average pooling. The result shows that ECA is able to put higher weights for the more informative channels in the appearance branch which helps the model understand the differences between the background and the human face.

\begin{figure}[ht]
\includegraphics[height=5cm]{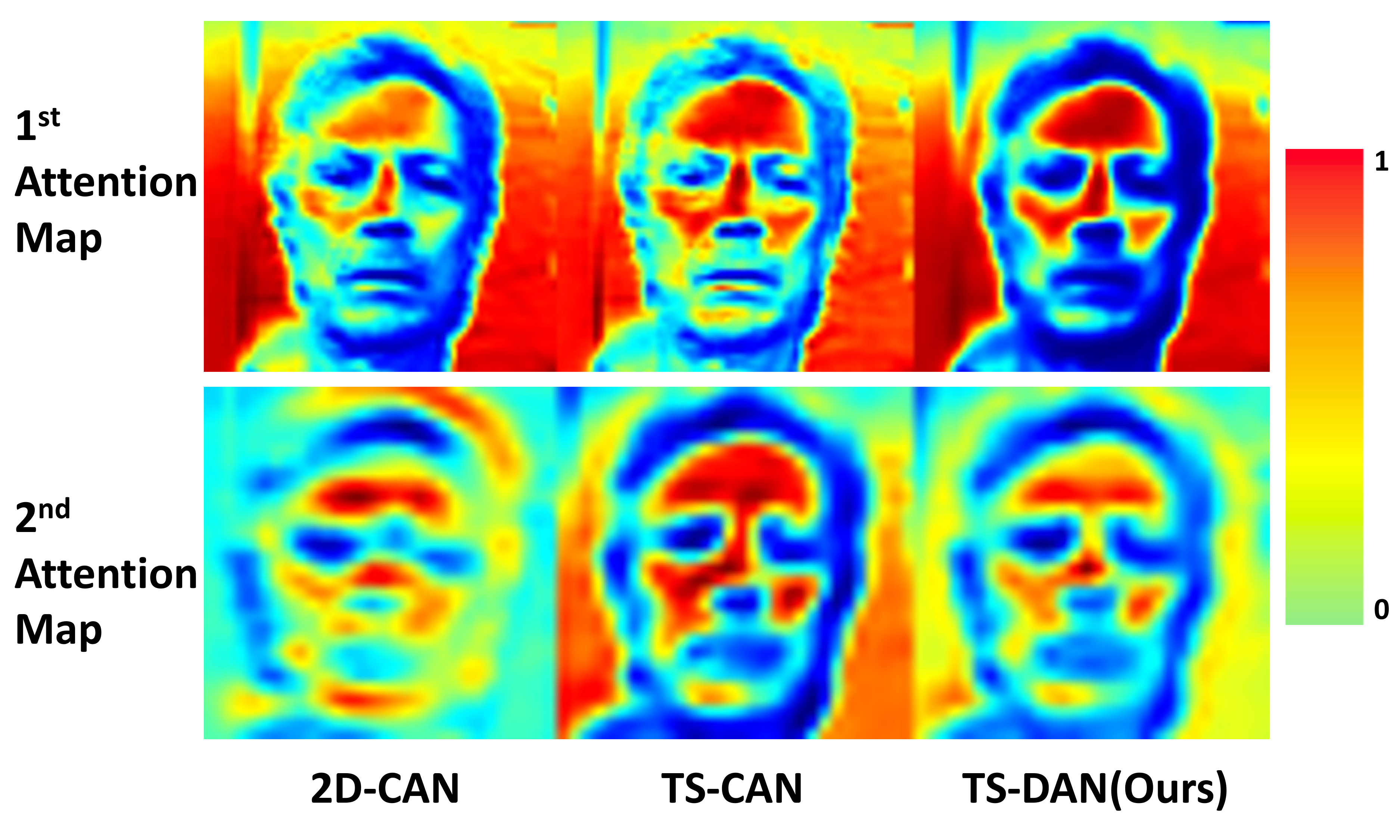}
\caption{Attention map visualization comparison from 2D-CAN\cite{chen2018deepphys}, TS-CAN\cite{liu2020multi}, TS-DAN (ours). The first attention map is after the second convolution layer and the second attention map is after the fourth convolution layer as shown in \figurename~\ref{architecture}. The first and second attention maps from our TS-DAN gives much better face skin localization and gives much lower weights for the background region.}
\label{attention_map}
\end{figure}

\figurename~\ref{attention_map} compares the first and second attention maps in 2D-CAN, TS-CAN and our proposed TS-DAN with 3 ECA layers. The first attention map is after the second convolution layer and the second attention map is after the fourth convolution layer as shown in \figurename~\ref{architecture}. Comparing the first attention map, 2D-CAN model shows higher weights only on a smaller face skin region, \textit(i.e.), the subject's left cheek and left forehead shows lower weights in soft attention map. TS-CAN model captures larger skin region however the boundary is blurry and shows false positive higher weights in eyelid region. The first attention map from our TS-DAN model clearly shows much better spatial localization on the skin region where physiological signal is stronger (forehead and cheeks). Furthermore, comparing the second attention map, attention map from TS-DAN shows larger contrast between face region and background with better boundary localization which indicates a better spatial and channel-wise feature extraction. TS-DAN gives lower weight on background than both 2D-CAN and TS-CAN. In summary, the attention map generated by TS-DAN gives higher weights on skin with better localization and gives lower weights on background, which helps improve network robustness and reduces background noise.

\subsection{Heart Rate and Respiratory Rate Estimation Evaluation}
Table.~\ref{cohface_single_task} compares heart rate and respiratory rate estimation accuracy between TS-CAN and TS-DAN in single task network. TS-DAN gives smaller MAE for full COHFACE dataset evaluation in both heart rate and respiratory rate estimation.


Table.~\ref{cohface} compares heart rate and respiratory rate estimation accuracy between TS-CAN and TS-DAN in multitask learning. MT-TS-DAN greatly decreases MAE, increases availability and SNR in heart rate estimation compared to MT-TS-CAN under full, clean, natural data evaluation. MT-TS-DAN achieves higher availability in respiratory rate estimation, however, slightly decreases respiratory rate MAE.

\begin{table*}[htbp]
\centering
\caption{Single task network for heart rate and respiratory rate estimation accuracy comparison using COHFACE dataset.}
\begin{tabular}{lccc|ccc|ccc}
\hline\noalign{\smallskip}
\bf{Method} & \multicolumn{3}{c}{\bf{Full}}  &  \multicolumn{3}{c}{\bf{Clean}} & \multicolumn{3}{c}{\bf{Natural}}\\
\noalign{\smallskip}
\hline
\noalign{\smallskip}
\textbf{Heart Rate} & MAE & Availability & SNR & MAE & Availability & SNR & MAE & Availability & SNR\\
TS-CAN\cite{liu2020multi} &1.751&0.907&6.712&1.771&  \textbf{0.963}& 7.099&\textbf{1.735}& 0.850& 6.325 \\
TS-DAN(ours) &\textbf{1.668}&\textbf{0.929}&\textbf{6.939}&\textbf{1.111}&0.939&\textbf{7.437}&2.224&\textbf{0.919}&\textbf{6.441} \\\hline
\textbf{Respiratory Rate} &  MAE & Availability & SNR & MAE & Availability & SNR & MAE & Availability & SNR\\
TS-CAN\cite{liu2020multi} &5.845&0.969&11.620&6.107&0.966&\textbf{11.300}&5.583&0.972&11.940 \\
TS-DAN(ours) &\textbf{5.350}&\textbf{0.979}&\textbf{12.010}&\textbf{5.380}&\textbf{0.979}&11.230&\textbf{5.320}&\textbf{0.979}&\textbf{12.790} \\
\hline
\end{tabular}
\label{cohface_single_task}
\end{table*}


\begin{table*}[htbp]
\centering
\caption{Multitask network for heart rate and respiratory rate estimation accuracy comparison using COHFACE dataset.}
\begin{tabular}{lccc|ccc|ccc}
\hline\noalign{\smallskip}
\bf{Method} & \multicolumn{3}{c}{\bf{Full}}  &  \multicolumn{3}{c}{\bf{Clean}} & \multicolumn{3}{c}{\bf{Natural}}\\
\noalign{\smallskip}
\hline
\noalign{\smallskip}
\textbf{Heart Rate} & MAE & Availability & SNR & MAE & Availability & SNR & MAE & Availability & SNR\\
MT-TS-CAN\cite{liu2020multi}  &4.265&0.619&4.454&3.236&0.688&5.190&5.295&0.550&3.718   \\
MT-TS-DAN(ours)&\textbf{1.807}&\textbf{0.885}&\textbf{6.651}&\textbf{1.281}&\textbf{0.888}&\textbf{7.312}&\textbf{2.333}&\textbf{0.883}&\textbf{5.990}\\
\hline
\textbf{Respiratory Rate} &  MAE & Availability & SNR & MAE & Availability & SNR & MAE & Availability & SNR\\
MT-TS-CAN\cite{liu2020multi} &\textbf{5.392}&0.976&\textbf{16.01}&\textbf{5.457}&0.965&\textbf{15.25}&\textbf{5.326}&0.986&\textbf{16.77}  \\ 
MT-TS-DAN(ours) &5.728&\textbf{0.991}&7.902&6.107&\textbf{0.994}&8.212&5.348&\textbf{0.988}&7.592\\
\hline
\end{tabular}

\label{cohface}
\end{table*}
\subsection{Cross-dataset Generalization}
To test whether our model can work well in different environment, such as different lightening conditions, different levels of head motions, we use our collected dataset to evaluate model generalization accuracy in heart rate estimation. Model trained with clean condition videos is tested using natural condition videos and vice versa. The model generalisation ablation study is shown in Table. \ref{generalization}. Our proposed model achieves much lower MAE in both generalization test.
\begin{table}[htbp]
\caption{Model generalization ablation. C2N and N2C in parentheses indicate training and testing data. C2N means training data is clean data and testing data is natural data. N2C means training data is natural data and testing data is clean data.}
\begin{center}
\centering
\begin{tabular}{lccc}\hline
  Model & MAE & Availability & SNR\\ \hline
2D-CAN \cite{chen2018deepphys} (C2N)    &4.082 & 0.681 & 4.058 \\  
TS-CAN \cite{liu2020multi}(C2N)    &3.230 & \textbf{0.847} & \textbf{5.499}\\  
TS-DAN (C2N)    &\textbf{2.000} & 0.708& 4.816\\ \hline
2D-CAN \cite{chen2018deepphys} (N2C)    &24.797 & 0.236 & 1.822 \\ 
TS-CAN \cite{liu2020multi}(N2C)    &3.617 & 0.699& 3.010 \\ 
TS-DAN (N2C)    &\textbf{1.091} & \textbf{0.805}& \textbf{3.772}\\ \hline
\end{tabular}
\end{center}
\label{generalization}
\end{table}
\section{Conclusions}
Temporal shift dual attention network (TS-DAN) was proposed to estimate heart rate and respiratory rate from a RGB video. We integrated both spatial attention and channel-wise attention to convolutional neural network architecture. The network can be applied in a single task or multitask learning fashion. It was demonstrated by experimental results that the proposed TS-DAN system offers the best accuracy on two benchmark datasets. 
 
{\small
\bibliographystyle{IEEEtran}
\bibliography{bib}
}
\end{document}